# Portfolio Choice with Indivisible and Illiquid Housing Assets: The Case of Spain

Sergio Mayordomo*, María Rodríguez-Moreno and Juan Ignacio Peña This paper studies the investment decision of the Spanish households using a unique data set, the Spanish Survey of Household Finance (EFF). We propose a theoretical model in which households, given a fixed investment in housing, allocate their net wealth across bank time deposits, stocks, and mortgage. Besides considering housing as an indivisible and illiquid asset that restricts the portfolio choice decision, we take into account the financial constraints that households face when they apply for external funding. For every representative household in the EFF we solve this theoretical problem and obtain the theoretically optimal portfolio that is compared with households' actual choices. We find that households significantly underinvest in stocks and deposits while the optimal and actual mortgage investments are alike. Considering the three types of financial assets at once, we find that the households headed by highly financially sophisticated, older, retired, richer, and unconstrained persons are the ones investing more efficiently.

*Sergio Mayordomo and María Rodriguez-Moreno are at the School of Economics and Business Administration, University of Navarra, Edificio Amigos, 31009 Pamplona (Spain), smayordomo@unav.es and mrodriguezm@unav.es . Juan Ignacio Peña is at the Department of Business Administration, Universidad Carlos III de Madrid, C/ Madrid 126, 28903 Getafe (Madrid, Spain), ypenya@eco.uc3m.es. We thank Anatoli Segura, Costanza Torricelli, Stefano Corradin, and participants in the IX INFINITI Conference, the 2011 Conference of the International Finance and Banking Society, and the XIX Meeting of the Spanish Finance Association for insightful comments. Mayordomo acknowledges financial support from MCI grant ECO2012-32554. Peña acknowledges financial support from MCI grant ECO2012-35023.

Corresponding author: Sergio Mayordomo, School of Economics and Business Administration, University of Navarra, Edificio Amigos, 31009 Pamplona (Spain), E-mail: smayordomo@unav.es. Phone: 0034 948 425600 Ext. 802435. Fax: + 34 948 425626

# 1. Introduction

This article analyzes the optimization problem faced by homeowners who must determine the amount of the investment in housing to be funded by a mortgage, which in turn conditions their demand for other financial assets. We frame this optimization problem as a reallocation problem in which the household's wealth is assigned across different assets: stocks, bank time deposits (our proxy for the risk-free asset), and mortgage, all of them conditioned to the housing value. The household's portfolio decision problem is subject to financial or borrowing restrictions due to its housing investment and includes liquidity shocks. Given that household portfolio choice literature suffers from a pervasive shortage of data on households' actual portfolio choices, our distinctive contribution is to employ the Spanish Survey of Household Finance (EFF). This unique data set enables us to combine a theoretical model with data of Spanish families' existent financial decisions and to obtain a more realistic assessment of the divergences between theory and practice.[1]

Housing is the main asset for households in many countries, particularly in Spain where it represented an average of 66.5% of the total assets in household portfolios in 2002 and 66.1% in 2005. According to the Spanish Household Budget Continuous Survey (ECPF), the percentage of homeowners was 84% in 2002 and 86% in 2005. These percentages suggest the importance of considering housing in analyses of households' optimal portfolio decisions. Initial studies in portfolio choice (Markowitz, 1952, Merton, 1971) did not include housing investments and more recent articles by Ameriks and Zeldes (2004) or Cocco, Gomes, and Maenhout (2005) focus solely on the impact of the life cycle on household portfolios. By excluding housing, these studies also fail to include housing-related financial liabilities (i.e., mortgages) in their models of households' portfolio decisions. One way to take housing investments into account is differentiating liquid from illiquid assets, as in Koren and Szeidl (2002), Schwartz and Tebaldi (2006) and Anglin and Gao (2011). This differentiation implicitly considers investments in housing and mortgages, though the high heterogeneity among illiquid assets likely makes it necessary to deal separately with housing and mortgage investments.

Some recent research includes housing and mortgage as additional financial assets or liabilities into household portfolio choice but they consider housing as a standard financial asset, in the sense that investors can decide the amount to invest in housing every period. For instance, Cocco (2005) employs a model in which every period the investor chooses the size of housing's share in an optimal portfolio. Other papers consider that housing is an indivisible, illiquid asset; that is, it can be thought of as a durable consumption good. In this line, Flavin and Yamashita (2002) focus on the impact of the portfolio constraint imposed by the consumption demand for housing on the household's optimal holdings of financial assets in the United States. An alternative view is presented in Kraft and Munk (2011) who derive closed-form solutions to life-cycle utility maximization including real estate. They are able to separate the consumption and investment decisions of housing by allowing for both renting the house and owning it. They also show that the wealth losses associated with inflexible housing decisions are very small.

---

[1] Other studies that use U.S. or U.K. micro data include Campbell and Cocco (2007), Cocco (2005), and Flavin and Yamashita (2002).

Based on these results we consider that housing may be viewed as a "restriction" on portfolio choice, as an investment already undertaken, that determines the investment in the other financial assets. When a household purchases a house, it does so because the house is consistent with either its true preferences or preferences restricted to its financial situation and so, its actual housing could differ from the desired one. Given that the investment in housing is considered as a constraint for the investment decision, we focus on those households living in their desired housing to reduce the risk that they would move or rent the house in the short-run. We estimate the desired house value using Mayordomo's (2008) method and consider housing as the desired one whenever its desired value is below the housing's market price. With this assumption, we posit that the desired housing represents an illiquid financial asset from the household's perspective that will not be traded in the near future. Moreover, this premise enables us to assume that households solve a problem with infinite time horizons in which housing appears as a permanent financial restriction. In addition, we focus on households who bought housing recently to reduce uncertainty about the preferences of the household and the probability of moving; under the assumption that the more years in a given housing, the higher is this uncertainty.

Our study presents methodological and empirical contributions to the existing literature. On the methodological side, we extend Koren and Szeidl's (2002) optimization methodology by including equity risk. Additionally, the model distinguishes among four classes of financial assets in the optimization problem (stocks, bank time deposits, mortgage, and housing) instead of analyzing just two asset categories (i.e., one liquid and one illiquid), as is commonly employed in prior literature. Additionally, the model includes the restrictions derived from the housing purchase that affect the mortgage decision and is applied on those households living in the desired housing to reduce the risk that they would move or rent the house in the short-run. Regarding the empirical contributions, we study the actual composition of the households' portfolios, noting different demographic and financial characteristics. Then, we estimate an optimal investment portfolio for individual households and for different groups of households attending to several demographic characteristics. Finally, we compare actual and optimal portfolios and study the demographic and financial characteristics explaining the deviations from the household's optimal portfolio, that is, deviations between actual and optimal investments.

Our baseline results are based on the fact that the average actual proportion invested by Spanish households in housing is 128.2% of their total net wealth. The actual proportions of total net wealth invested in stocks, bank time deposits, and mortgage are 0.6%, 3.1%, and -31.9%, respectively; while the theoretically optimal proportions are 1.6%, 4.9%, and -34.7%, respectively. We find a positive relationship between the optimal proportion invested in stocks and the households' education level and net wealth. The proportion of housing and the optimal share of mortgage decreases with age, education level, and household net wealth. Finally, the optimal amount to be invested in bank time deposits is around 4-5% and nearly constant across demographics groups.

Statistical tests indicate that Spanish households tend to invest significantly less in stocks and deposits than theory indicates they should. Nevertheless, the optimal and actual proportions invested in mortgages are statistically indistinguishable in all the households' categories. Considering the three types of financial assets at once, we find that the households headed by highly financially sophisticated, older, retired, richer, and unconstrained persons are the ones investing more efficiently. The remaining categories of households can improve their liquidity and capacity to face liquidity shocks if they slightly increase their share of mortgage and invest this extra liquidity in equity and/or bank-time deposits.

From our study of portfolio rebalancing from 2002 to 2005 we find that, on average, households under-invest in stocks in similar proportions in 2002 and 2005 and slightly less so in deposits. Actual investment in mortgage is similar in both periods and very close to the optimal one. Finally, our analysis of the investments decision of households that purchased their hosing close to the two survey dates shows that the average optimal investment in mortgage is slightly higher in 2005 than in 2002. These results could be due to the increase in housing prices and to the somewhat less restrictive banking practices around 2005. The optimal investment in bank deposits is similar but the optimal investment in stocks in 2005 is much higher than in 2002.

The rest of the paper is organized as follows. In Section 2, we describe the theoretical model, the optimization technique, and the calibration procedure. In Section 3, we describe the data. Section 4 presents the empirical results. In Section 5, we present a comparative statics analysis and Section 6 concludes.

## 2. Model

We present the portfolio decision problem for a household subject to financial or borrowing restrictions due to its housing investment and exposure to liquidity shocks. As in Koren and Szeidl (2002), we use a relatively simple functional form that enables us to find a semi-analytic solution for the value function and for the optimal consumption and portfolio policy. We analyze a time-discrete portfolio choice decision model for infinitely persisting households with exogenous initial wealth and labor income, such that the utility function has a constant relative risk aversion coefficient.

Assuming infinite horizon is a crucial assumption that impedes to properly address life-cycle implications in this setting.[2]

---

[2] We refer to Kraft and Munk (2011) for a full account of the impact of housing investment decisions on the life-cycle optimal portfolio choice.

Nevertheless, this is not the aim of this paper and this assumption facilitates the optimization procedure because it eliminates the dependence of the Bellman equation on time. This lack of dependence allows us to solve the optimization problem for a steady-state. Moreover, as Brandt (2009) states one would expect the sequence of solutions to a finite horizon problem to converge to that of the corresponding infinite horizon problem as the horizon increases. In the case of CRRA utility and empirically sensible return processes, this convergence appears to be quite fast. Brandt (1999), Barberis (2000), and Wachter (2002) all document that 10- to 15-year constant relative risk aversion (CRRA) portfolio policies, as the one employed in our study, are very similar to their infinite horizon counterparts. Brandt (1999) argues that this rapid convergence suggests that the solution to the infinite horizon problem can in many cases be confidently used to study the properties of long- but finite-horizon portfolio choice in general (e.g., Campbell and Viceira (1999,2002)).[3]

## 2.1. The Assets

We assume that the household can invest in three types of assets: illiquid, semi-liquid, and liquid. The set of illiquid assets comprises the house and its mortgage. We assume that among the illiquid assets, the only tradable asset (although not in the very short-run) is the mortgage; whereas the investment in housing is given and remains constant over time which implies a constrained investment strategy. As shown in Kraft and Munk (2011) the expected utility of a constrained strategy is smaller than the expected utility of the optimal unconstrained strategy. However, the economic significance of this wealth loss is small. Therefore, we restrict our analysis to households living in their desired house (see Appendix A.2 for details about its estimation), to reduce the risk of moving in the short-run.[4] Choosing households that dwell in their desired home is crucial; otherwise, there is uncertainty about the time period the household will remain in the less desired home, such that the house provides "transitory" housing to be sold before buying the desired one. The way housing affects allocation choices is different if the level of consumption of housing services is optimally set or if it is considered as an exogenous constraint affecting the investment decision. Focusing in households living in the desired housing guarantees the role of housing not just as an investment but also a durable consumption good. Thus, independently on whether the housing investment is the optimal or not, the households who live in their desired home should not be planning to optimize the investment in housing because they are not interested in moving. For this reason, we can consider this investment as an investment already undertaken.

---

[3] Cultural reasons may also support the assumption on infinite horizon decisions of Spanish households. Reher (1998) considers Spain as a "strong family country" where families play a protective function by caring about intergenerational relationships. The family structure is more extended, the co-residence between parents and children is frequent and the economic help is prolonged. Thus, the type of problem presented in our analysis does make intuitive sense given that the households care about the consumption levels of not just themselves but also their heirs and so, they might be planning their consumption over an infinite horizon.

[4] We ignore contractual savings for housing, because the decision about the amount to allocate to this asset occurs before the housing purchase, which we consider a given, so this asset should not affect the portfolio choice or reallocation problem.

As a semi-liquid asset we include bank time deposits that yield a riskless return. We employ deposits instead of other fixed-income instruments such as short- and long-run Government bonds because of the scarce presence of these assets in Spanish households' portfolios. In fact, according to the EFF less than 2% of households invest in either sovereign or corporate fixed-income instruments. In any case, the correlation between the annual returns of deposits and short-term Government bonds from 1992 to 2002 was 0.97, which suggests that they are almost perfect substitutes. In this setting, bank time deposits are considered as the risk free asset because of, among other reasons, the existence of the deposit insurance scheme at the time of the EFF survey waves (2002 and 2005).[5] Bank time deposits commonly impose restrictions on the investment horizon or the availability of funds to make payments. Moreover, this asset encompasses a wide variety of investment products with varying characteristics. To be consistent with the information provided by the EFF, we employ a standard bank time deposit that does not have any specific maturity. Such deposits are not available to households in the very short run, unless they pay a preset cancellation fee. Therefore, the household employs these deposits in the very short run, subject to a known penalty payment, or waits until maturity. The latter decision implies that these deposits may be employed to make payments or consumed in the future. In this last case, the household receives the interest payment and does not pay any cancellation fee.[6]

The liquid asset, which is also the riskier, is a stock portfolio that can be employed in the very short run for consumption.[7] We consider three states of nature (scenarios) determined by the stock portfolio return's behavior that characterizes the state of the economy. As the reference stock portfolio we employ the IBEX 35 stock index. This selection is motivated by the home bias phenomenon (French and Poterba, 1991), according to which investors tend to over-invest in domestic stocks. Recently, Jochem and Volz (2011) studied the portfolio holdings in the euro area countries showing that the home bias phenomenon in Spain is pervasive (the highest in the euro area), being close to 100% in 1991, around 90% in 1998, and around 85% in 2008. Similar results have been also found by Sørensen et al. (2005) and Sercu and Vanpée (2012), among others.[8]

---

[5] Before 2008 the deposit insurance scheme covered up to a maximum of €20,000 per depositor and bank while according to Panel B of Table 1 Spanish households held on average €5,035 on deposits in 2002. In fact, this scheme was increased in 2008 to cover up to a maximum of €100,000 per depositor and bank.
[6] The standard bank time deposit is similar to a fixed-term deposit.
[7] Mutual funds might be additional instrument to consider in this optimization problem. However, most mutual funds' investments involve equity or fixed-income, so we consider this financial instrument partly represented by deposits and equity instruments. The same argument holds for the absence of pension funds among the set of assets we consider. Thus, we avoid using redundant assets but still select the most representative assets of Spanish households' portfolio.
[8] Possibly because of this home bias behavior the EFF does not contain information on whether the shares held by Spanish households were issued by domestic or foreign institution. In spite of that, as an aim to check the robustness of our analysis we have used the EUROSTOXX50 as the benchmark to define the stock returns and find similar results given that the correlation between the EUROSTOXX50 and IBEX35 annual returns is around 80%.

We therefore assume a favorable ($H$) state when the average real annual return of the Spanish stock index (including dividends) exceeds the 75th percentile of its distribution in 2002 (37.77%). So, in this scenario the stock portfolio return is equal to 37.77% and it occurs with probability 25%, which is represented by the parameter $\lambda_1$. The intermediate scenario ($M$) occurs with a probability of 50% (i.e., $\lambda_2 = 0.5$) when the average real return falls between the 25th (-13.87%) and 75th (37.77%) percentiles. In an intermediate scenario, the investor receives the average of the stock's returns distribution that is equal to 16.86%. Finally, the unfavorable scenario (L) occurs with a probability of 25% (i.e., $1 - \lambda_1 - \lambda_2$) and yields an average real annual return equal to -13.87%, which corresponds to the 25th percentile.

## 2.2. Model Timing

The timing for asset trade, consumption, and interest earned is illustrated in Figure 1. We consider that the market for less liquid assets (mortgages and deposits, assuming the household does not pay the deposit cancellation fee) operates with a lag with respect to the liquid asset's market. For the sake of clarity, we describe the timing between *t* and *t+1* as well as the investment and events that took place at *t-1* affecting period *t*. At the beginning of period *t*, the interest ($R_t$) earned for holding the portfolio between *t* and *t+1* is paid in advance. The composition of this portfolio was set up at *t-1* before the state of the nature affecting period *t* and the potential liquidity shock are realized. There are three possible states of nature: good, intermediate and bad which determine the state of the economy and hence, the return of the investment, $R_t$. Then, a liquidity shock which increases the marginal utility of consumption is realized and accordingly, the household chooses consumption ($C_t$) subject to its current liquidity constraints. At the same time (end of period *t*) the household decides the composition of the portfolio to be held for the next period *t+1*.

The households have the ability to decide on consumption given that they can sell the liquid and semi-liquid assets that were purchased at *t-1* (order 1 in Figure 1) and consume them immediately. For this reason, part of the consumption at time *t* is determined by the investment at *t-1* and so, by the state of nature that is known at the end of *t-1*. On the contrary, illiquid assets are not available for immediate consumption and its portfolio share cannot be modified until the next buy/sell order is placed. This formulation captures the time needed to sell a given asset, which reflects its liquidity.

The state of nature at period t affects the returns obtained at the beginning of *t*, and the consumption at *t* ($C_t$). Additionally, a liquidity shock occurring at time *t* also affects $C_t$ given that the household can liquidate the liquid and semi-liquid assets to be consumed. This creates uncertainty when the investment is done about the potential state of nature and the probability of suffering a liquidity shock. In summary we have six separate consumption values depending on the three states of nature and also depending on the materialization of a liquidity shock or otherwise. Both the liquidity shock and the state of nature are assumed to be independent over time, in any of the six possible regime/state combinations.

## 2.3. Optimization Problem

The household maximizes the following utility function:

$$\max_{\{C_t, \alpha_{t+1}\}} E_0 \sum_{t=0}^{\infty} \beta^t \chi_t^\theta \frac{C_t^{1-\theta}}{1-\theta} \qquad (1)$$

where $C_t$ represents consumption at time $t$, $\theta$ is the relative risk aversion coefficient, $\beta$ is the subjective discount factor that measures the preference for future consumption, and $\chi_t$ is the taste or liquidity shock at a given time $t$, independent and identically distributed over time. We assume that when a household suffers a liquidity shock in a given period, the marginal utility of consumption increases with respect to that observed in normal times, which causes households to consume more during a liquidity shock. This shock can take two only values:

$$\chi_t = \begin{cases} \gamma > 1 \text{ with probability } \mu \\ 1 \text{ with probability } 1 - \mu. \end{cases} \qquad (2)$$

When $\chi_t = \gamma$, the investor suffers a liquidity shock in period $t$. If $\gamma > 1$, the marginal utility of consumption is higher during a liquidity shock than at normal times, which leads households to consume more during a liquidity shock.

The households are subjected to a series of restrictions that are summarized, jointly with the objective function, in the following equation:

$$\max_{\{C_t, \alpha_{t+1}\}} E_0 \sum_{t=0}^{\infty} \beta^t \chi_t^\theta \frac{C_t^{1-\theta}}{1-\theta} \qquad (3)$$

$$\text{s.t.} : \quad W_t = \alpha_t'[W_t] \qquad (\text{i})$$

$$\alpha_t'[1] = 1 \qquad (\text{ii})$$

$$\alpha_{t,M} \leq 0 \qquad (\text{iii})$$

$$\alpha_{t,i} \geq 0 \quad \text{for } i = 1, \ldots, 3 \qquad (\text{iv})$$

$$R_{t,p} = 1 + r_{t,p} \qquad (\text{v})$$

$$\alpha_t' r_t = r_{t,p} \qquad (\text{vi})$$

$$C_t \leq \left(\alpha_{t,1} + \alpha_{t,2}(1 - fee) + r_{t,p} - \alpha_{t,2} r_{t,2}\right) W_t + L_t \qquad (\text{vii})$$

$$W_{t+1} = R_{t,p} W_t - C_t + L_t \qquad (\text{viii})$$

$$-0.8 H_t \leq \alpha_{t,M} W_t \leq 0 \qquad (\text{ix})$$

$$R_{t,M}\alpha_{t,M}W_t \leq 0.33Y_t \quad\quad\quad (x)$$

where $W_t$ represents the total financial net wealth obtained as the sum of the wealth invested in liquid, illiquid, and semi-liquid assets $\alpha_t'[W_t]$ (restriction *i*). The vector $\alpha_t'$ contains the share of wealth held in liquid (stocks, $\alpha_{t,1}$) semi-liquid (bank time deposits, $\alpha_{t,2}$), and illiquid (housing, $\alpha_{t,3}$, and mortgage, $\alpha_{t,M}$) assets, respectively; at the beginning of *t* such that the sum of the components of the vector is equal to 1 (restriction *ii*). The proportion of wealth invested in the mortgage, $\alpha_{t,M}$, which is equivalent to a short position in a bond, is set to be negative (restriction *iii*).[9] The other portfolio weights are restricted to be positive (restriction *iv*), because we assume short sales are not allowed. The $R_t$ and $r_t$ vectors contain the gross and net real return, respectively, of all financial assets (deflated by the appropriate price index) from *t* to *t+1* that are obtained in advance at the beginning of period *t* (restriction *v*). In contrast, $r_{t,p}$ refers to the net portfolio return, composed of the weighted average of $r_t$, where the weights depend on the portion of wealth invested in the corresponding asset (restriction *vi*).

Because household needs to borrow to fund the housing purchase, consumption cannot be higher than total holdings in stocks and deposits plus the return earned on asset holdings and the labor income at year *t*, which we denote as $L_t$.[10] Moreover, when a household withdraws the bank time deposit before maturity, they pay a cancellation fee, denoted $fee$, and do not receive the returns derived from such investment ($\alpha_{t,2}r_{t,2}$).[11] Restriction (*vii*) illustrates this scenario. One step ahead total net wealth ($W_{t+1}$) is the updated portfolio value plus labor income minus consumption in period *t* (restriction *viii*).

The model also imposes two constraints related to standard mortgage policy recommendations. First, the mortgage must be lower than 80% of the housing's value at time *t* ($H_t$). The value 0.8 is common for all the households and compatible with a bank's standard provisions.[12] This constraint aligns with good banking practices in banks, which generally do not lend the whole value of the house, to avoid moral hazard problems and to ensure the compatibility of incentives (restriction *ix*). Second, the mortgage payments must be lower than 33% of household income at time $t$ ($Y_t$). This financial constraint, set by financial institutions, helps ensure that the household will be able to pay the mortgage and avoids potentially high delinquency ratios. The value 0.33 represents an approximation in relation to the income requirements that a given household must fulfill to obtain the loan (restriction *x*).

---

[9] Other illiquid financial assets such as bonds or long-run deposits would be a perfect substitute for an opposite position in the mortgage. In that case, if the illiquid asset's returns are higher than the mortgage returns, the investor would demand the maximum amount of mortgage to invest in that asset. If the returns are lower than the mortgage rate, investors will not invest in it. Therefore, the shares invested in the illiquid asset and the mortgage cancel out each other.

[10] We consider labor income an implicit holding of safe assets, such that the household receives a fixed amount of money every month for the rest of its life.

[11] This fee is equal to 5%, the average bank time deposits' return during the study period, as we explain in the calibration section. We use a fee equal to the previous percentage because if the bank time deposit is cancelled, the investor does not receive the whole notional but rather the notional minus a given amount, which cannot exceed the received interest. Because we work in annual terms, we use this 5% to control for the received interest without imposing any concrete investment horizon.

[12] If this proportion is higher than 0.8, the provisions offered by the lender must increase. In Flavin and Yamashita (2002), the households should maintain a portfolio in which the amount of their mortgage is equal to the value of their house.

The liquidity shock is independently distributed over time, so there are only two state variables, total net wealth ($W_t$) and the share of financial assets ($\alpha_t$). With these two states in terms of the shock realization, we consider two different types of consumption for each household. When a liquidity shock materializes, the household consume $C_{1,t}$, otherwise, they consume $C_{2,t}$. In addition to the liquidity shock state, we assume there are three states of nature that are closely related to the state of the economy. Ideally, we would consider three different states of nature for each asset (i.e., 64 states), but for the sake of tractability we concentrate on three states of nature defined by the stock index due to the close relation between the economy and the stock index and the high volatility of this asset compared to the others. So, we include six control variables for consumption, depending on the states of nature associated with both the liquidity regime and the economic regime. Moreover, we have an extra control variable, that is, the next period's portfolio share for every financial asset.[13]

The value function and household's optimal program can be characterized by the following Bellman equation:

$$
\begin{aligned}
&V(W, \alpha') \\
&= \max_{\alpha'} \Bigg\{ \mu \Bigg[ \begin{array}{l} \lambda_1 \left( \max_{C_1^H \leq Liq.Wealth^H} \left[ \gamma^\theta \frac{C_1^{H,1-\theta}}{1-\theta} + \beta V(R_p^H W - C_1^H + L, \alpha') \right] \right) \\ + \lambda_2 \left( \max_{C_1^M \leq Liq.Wealth^M} \left[ \gamma^\theta \frac{C_1^{M,1-\theta}}{1-\theta} + \beta V(R_p^M W - C_1^M + L, \alpha') \right] \right) \\ + (1-\lambda_1-\lambda_2) \left( \max_{C_1^L \leq Liq.Wealth^L} \left[ \gamma^\theta \frac{C_1^{L,1-\theta}}{1-\theta} + \beta V(R_p^L W - C_1^L + L, \alpha') \right] \right) \end{array} \Bigg] \\
&\quad + (1-\mu) \Bigg[ \begin{array}{l} \lambda_1 \left( \max_{C_2^H \leq Liq.Wealth^H} \left[ \frac{C_2^{H,1-\theta}}{1-\theta} + \beta V(R_p^H W - C_2^H + L, \alpha') \right] \right) \\ + \lambda_2 \left( \max_{C_2^M \leq Liq.Wealth^M} \left[ \frac{C_2^{M,1-\theta}}{1-\theta} + \beta V(R_p^M W - C_2^M + L, \alpha') \right] \right) \\ + (1-\lambda_1-\lambda_2) \left( \max_{C_2^L \leq Liq.Wealth^L} \left[ \frac{C_2^{L,1-\theta}}{1-\theta} + \beta V(R_p^L W - C_2^L + L, \alpha') \right] \right) \end{array} \Bigg] \Bigg\}
\end{aligned}
$$

(4)

where $Liq.Wealth^i = (\alpha_1 + \alpha_2(1-fee) + r_p^i - \alpha_2 r_2)W + L$ (for $i = H, M, L$) refers to the liquid wealth in each state of nature, as summarized in restriction *vii* of Equation (3) in the optimization problem. In turn, $C_1^H$, $C_1^M$, and $C_1^L$ represent the consumption profile in a favorable (H), an intermediate (M), and a unfavorable (L) state if a liquidity shock occurs. The consumption profile in a normal regime with no liquidity shock is defined by $C_2^H$, $C_2^M$, and $C_2^L$ for the same three states: favorable, intermediate, and unfavorable, respectively. Parameter $\mu$ represents the probability of the occurrence of a liquidity shock. Parameters $\lambda_1$ and $\lambda_2$ represent the probabilities of the occurrences of a favorable and an intermediate state, respectively. The returns $r_p^H$, $r_p^M$, and $r_p^L$ ($R_p^H$, $R_p^M$, and $R_p^L$) refer to the net (gross) portfolio returns, obtained in the three possible states, H, M, and L, respectively. Finally, the vector $\alpha'$ denotes the fraction of wealth in each of the financial assets after this period's order is executed.

---

[13] Koren and Szeidl (2002) do not consider the states of nature defined by the asset returns and limit the investor's uncertainty to the existence of a liquidity shock that exists in two potential consumption regimes. Moreover, they only consider two assets, liquid and illiquid, without taking into account any division in each category.

The intuition behind the Bellman equation is as follows. A liquidity shock may happen with probability $\mu$. When a liquidity shock materializes, current consumption delivers higher marginal utility, so the marginal utility increases by $\gamma^\theta$. However, consumption cannot be more than current liquid wealth, and the constraint is binding in the maximization problem (Equation (4)). The next-period net wealth equals the net wealth plus the interest earned in this period and the labor income minus the consumption. The second part of the maximization problem (Equation (4)) describes the case in which there is no liquidity shock, but we still must consider the effect of the state of the economy. This maximization can be interpreted similarly to that which corresponds to the part of the liquidity shock. We assume that both the liquidity shock and the state of nature are independent over time, in any of the six possible regime/state combinations, and the household chooses the same optimal portfolio share for the next period, $\alpha'$, before the state of nature is known.

## 2.4. Optimization Methodology

Because all the constraints are linear in consumption and wealth, and the utility function is a power utility, the value function is homogeneous to degree $1 - \theta$ in wealth. In turn we can show that there exists a function $\phi(\alpha)$ such that:[14]

$$V(W, \alpha') = \phi(\alpha)^{-\theta} \frac{W^{1-\theta}}{1 - \theta}. \qquad (5)$$

The existence of homogeneity in wealth means that the investor's portfolio choice for next period is independent of wealth. The implication is that there is an optimal portfolio share to be invested in the assets, $\alpha_i^{*\prime}$ for $i$ in 1,2,3, and $M$. Thus, according to the previous expression, the value function is maximal for any wealth level when $\phi(\alpha)^{-\theta}\frac{1}{1-\theta}$ is maximal. For $\theta > 1$, it can be formally expressed as

$$\alpha^* = \arg\max_{\alpha} \phi(\alpha). \qquad (6)$$

---

[14] Koren and Szeidl (2002) offer a proof for this statement; they consider a given positive constant $k$, such that $k > 0$. Because all constraints are linear, the optimal contingent plan for initial wealth $kW$ and portfolio $\alpha$ will be just $k$ times the optimal contingent plan for initial wealth $W$ and portfolio share $\alpha$. Then, the per period utility is homogeneous to the degree $1 - \theta$, so it follows that the total value of initial wealth $kW$ will be equal to $k^{1-\theta}$ times the total value of initial wealth $W$. Therefore the value function of the problem will also be homogeneous to the degree $1 - \theta$, and it is also a smooth function of $W$. For this reason, it must assume the functional form in Equation (5).

To solve this optimization problem, we adopt Koren and Szeidl's (2002) methodology. We already know that the optimal policy involves a constant share of liquid wealth and that the value function is separable in $W$. Thus, we begin by defining the subset of all feasible policies from all restrictions that involve portfolio shares $\alpha$. The optimal policy falls within this subset and corresponds to the one that offers higher utility for the consumer (i.e., first best policy). Liquidity shocks could appear during households' lives, causing the household to spend all the liquid wealth on consumption, because before the next period, they have a new opportunity to rebalance their portfolios. These possible rebalances make holding more liquidity than needed during a liquidity shock unnecessary. However, consumption during a liquidity shock is limited by the amount of liquid assets and the total portfolio's returns, such that the consumption in this regime is less than the consumption in the unconstrained first best policy. In an unfavorable state, the household may be unwilling to hold stocks, because they give negative returns, but we assume that the preference for consumption in the case of a liquidity shock is so high that the household sell their stocks in spite of any potential losses. Similarly, a preference for consumption leads the household to withdraw their bank time deposits, despite any cancellation fees.

For an optimal policy of this form, we focus on a set of policies consistent with the optimal policy. That is, during a liquidity shock, all the liquid and semi-liquid wealth is consumed; otherwise, the households choose consumption optimally. We solve the problem for fixed values of $\alpha$, denoted $\tilde{\alpha}$, which include the optimal policy, and impose the extra restrictions we detailed previously. This approach is equivalent to solving the optimal problem with one extra restriction, which eliminates some of the control variables such as consumption given a liquidity shock. This simplification leads to a modified problem that is easier to solve.

Accordingly, we set the grid of $\tilde{\alpha}$ that satisfies all restrictions and employ them to obtain the optimal value given by the portfolio shares $\tilde{\alpha}^*$. Thus, we first must characterize the value function of the modified problem for any $\tilde{\alpha}$, then maximize by substituting over the grid of $\tilde{\alpha}$ to obtain $\tilde{\alpha}^*$. This value function is denoted $\tilde{V}(W, \tilde{\alpha})$ and we can show that the value of the modified problem is less than or equal to the value of the original problem, with equality, if and only if $\tilde{\alpha}^* = \alpha^*$:

$$\tilde{V}(W, \tilde{\alpha}') \leq V(W, \tilde{\alpha}'). \qquad (7)$$

The homogeneity property of the utility function means the value function of the modified problem also will be homogeneous to the degree $1 - \theta$ in wealth. As in Equation (5), we have:

$$\tilde{V}(W, \tilde{\alpha}') = f(\tilde{\alpha})^{-\theta} \frac{W^{1-\theta}}{1-\theta} \tag{8}$$

In summary, we recast the initial optimization problem into a simpler modified problem, and to solve it, we set a grid of $\tilde{\alpha}$ that represents different portfolios that hold all required restrictions. Once the grid of portfolios is set, we replace each portfolio in the value function $\tilde{V}(W, \tilde{\alpha}')$ to find the portfolio that gives the maximum $\tilde{V}(W, \tilde{\alpha}')$. From Equation (8), to find the optimal value function, we must define the function $f(.)$ first. Once we define $f(.)$, the problem is reduced to maximizing $f(\tilde{\alpha})$ in $\tilde{\alpha}$ to find $\tilde{\alpha}^*$ and to finding the optimal consumption when there is no liquidity shock, that is, the first best rule consumption.

The value function under the modified problem is given by the following Bellman equation:

$$\tilde{V}(W, \tilde{\alpha}') = \begin{cases} \mu \begin{bmatrix} \lambda_1 \left[ \gamma^\theta \frac{C_1^{H,1-\theta}}{1-\theta} + \beta \tilde{V}(R_p^H W - C_1^H + L, \tilde{\alpha}') \right] \\ +\lambda_2 \left[ \gamma^\theta \frac{C_1^{M,1-\theta}}{1-\theta} + \beta \tilde{V}(R_p^M W - C_1^M + L, \tilde{\alpha}') \right] \\ +(1-\lambda_1-\lambda_2) \left[ \gamma^\theta \frac{C_1^{L,1-\theta}}{1-\theta} + \beta \tilde{V}(R_p^L W - C_1^L + L, \tilde{\alpha}') \right] \end{bmatrix} \\ + (1-\mu) \max_{\tilde{\alpha}'} \begin{bmatrix} \lambda_1 \left( \max_{C_2^H \leq Liq.Wealth^H} \left[ \frac{C_2^{H,1-\theta}}{1-\theta} + \beta V(R_p^H W - C_2^H + L, \alpha') \right] \right) \\ +\lambda_2 \left( \max_{C_2^M \leq Liq.Wealth^M} \left[ \frac{C_2^{M,1-\theta}}{1-\theta} + \beta V(R_p^M W - C_2^M + L, \alpha') \right] \right) \\ +(1-\lambda_1-\lambda_2) \left( \max_{C_2^L \leq Liq.Wealth^L} \left[ \frac{C_2^{L,1-\theta}}{1-\theta} + \beta V(R_p^L W - C_2^L + L, \alpha') \right] \right) \end{bmatrix} \end{cases} \tag{9}$$

where by assumption, $C_1^H = (\alpha_1 + \alpha_2(1-fee) + r_p^H - \alpha_2 r_2)W + L = Liq.Wealth^H$, $C_1^M = (\alpha_1 + \alpha_2(1-fee) + r_p^M - \alpha_2 r_2)W + L = Liq.Wealth^M$ and $C_1^L = (\alpha_1 + \alpha_2(1-fee) + r_p^L - \alpha_2 r_2)W + L = Liq.Wealth^L$. This expression is obtained by rewriting the original Bellman equation, imposing the restrictions that the portfolio share must be equal to $\tilde{\alpha}$, and assuming that the household consumes all liquid wealth (stocks, deposits, assets' returns, and labor income) during a shock. The last restriction implies that in the modified problem, only consumption in the normal state ($C_2^H, C_2^M$, and $C_2^L$) remains as a choice variable. This consumption level is determined as follows: if the household's liquidity constraint is not binding in the normal state (because it has enough cash), then $C_2^i$ (for $i = H, M, L$) can be chosen according to the first-order condition of the problem.

From the Bellman Equation (9), and using the functional form of Equation (8), we derive the first-order condition of $C_2^i$:

$$C_{foc,2}^i = \frac{f(\tilde{\alpha})\beta^{-1/\theta}(R_p^i W + L)}{1 + \beta^{-1/\theta} f(\tilde{\alpha})} \text{ for } i = H, M, L. \tag{10}$$

If the liquidity constraint binds and the household consumes all its liquid wealth, then:

$$C^i_{const,2} = (\alpha_1 + \alpha_2(1-fee) + r^i_p - \alpha_2 r_2)W + L \; for \; i = H, M, L \qquad (11)$$

Note that $C^i_{foc,2}$ is only feasible if $C^i_{foc,2} \leq C^i_{const,2}$. For this reason, the household chooses $C^i_2$ as the minimum of these two consumption levels:

$$C^i_2 = \min\left\{(\alpha_1 + \alpha_2(1-fee) + r^i_p - \alpha_2 r_2)W + L, \frac{f(\tilde{\alpha})\beta^{-1/\theta}(R^i_p W + L)}{1+\beta^{-1/\theta}f(\tilde{\alpha})}\right\} \qquad (12)$$

According to these expressions, there should be a level of the portfolio weights at which the liquidity constraint becomes binding in the normal state, and hence, $C^i_{foc,2}$ and $C^i_{const,2}$ coincide. For such a level of portfolio weights in which the consumption obtained from the first-order condition is higher than the constraint consumption in all states of nature, the corresponding value function, $f_1(\tilde{\alpha})$, is given by

$$f_1(\tilde{\alpha}) = \left[\frac{1-\beta(1-\tilde{\alpha}_1-\tilde{\alpha}_2(1-fee))^{1-\theta}}{\mu\gamma^\theta+(1-\mu)}\right]^{\frac{1}{\theta}}\left[\lambda_1\left(\tilde{\alpha}_1+\tilde{\alpha}_2(1-fee)+r^H_p\right.\right.$$
$$\left.-\tilde{\alpha}_2 r_2+\frac{L}{W}\right)^{1-\theta}+\lambda_2\left(\tilde{\alpha}_1+\tilde{\alpha}_2(1-fee)+r^M_p-\tilde{\alpha}_2 r_2+\frac{L}{W}\right)^{1-\theta} \qquad (13)$$
$$\left.+(1-\lambda_1-\lambda_2)\left(\tilde{\alpha}_1+\tilde{\alpha}_2(1-fee)+r^L_p-\tilde{\alpha}_2 r_2+\frac{L}{W}\right)^{1-\theta}\right]^{\frac{-1}{\theta}}.$$

The value function's implicit expression for the level of portfolio weights in which the consumption obtained from the first-order condition is lower than or equal to the constraint consumption in all states of nature, $f_2(\tilde{\alpha})$, is obtained as the root of the following equation:

$$1 = \left[(1-\tilde{\alpha}_1-\tilde{\alpha}_2(1-fee))^{1-\theta}\beta\mu\right]$$
$$+f_2(\tilde{\alpha})^\theta\gamma^\theta\mu\left[\lambda_1\left(\tilde{\alpha}_1+\tilde{\alpha}_2(1-fee)+r^H_p-\tilde{\alpha}_2 r_2+\frac{L}{W}\right)^{1-\theta}\right.$$
$$+\lambda_2\left(\tilde{\alpha}_1+\tilde{\alpha}_2(1-fee)+r^M_p-\tilde{\alpha}_2 r_2+\frac{L}{W}\right)^{1-\theta} \qquad (14)$$
$$\left.+(1-\lambda_1-\lambda_2)\left(\tilde{\alpha}_1+\tilde{\alpha}_2(1-fee)+r^L_p-\tilde{\alpha}_2 r_2+\frac{L}{W}\right)^{1-\theta}\right]$$
$$+(1+\mu)\beta\left[\lambda_1 R^{H1-\theta}_p+\lambda_2 R^{M1-\theta}_p+(1-\lambda_1-\lambda_2)R^{L1-\theta}_p\right]\left(1+\beta^{-\frac{1}{\theta}}f_2(\tilde{\alpha})\right)^\theta.$$

Appendix A.1 contains the implicit expressions for the remaining value functions, which depend on the scenario in which the liquidity constraint is binding. We obtain the corresponding $f_j(\tilde{\alpha})$ for each value of $\tilde{\alpha}$, then find the optimal portfolio weights that lead to the maximum $f_j(\tilde{\alpha})$. The numerical strategy that we use to solve the optimization problem is similar to that employed by Koren and Szeidl (2002). The grid of vectors $\alpha s$ represent different portfolio weights, so for each candidate portfolio weight $\tilde{\alpha}$, we solve $f_2(\tilde{\alpha})$, which implies that the liquidity constraint is not binding in any of the three scenarios (i.e., $i = H, M, L$). With the value of $f_2(\tilde{\alpha})$, we can verify whether the assumption that the liquidity constraint is not binding in any of the three scenarios holds. We thus test if the corresponding consumption levels obtained under $f_2(\tilde{\alpha})$ in the three states of nature verify that $C_{foc,2}^H \leq C_{const,2}^H \cap C_{foc,2}^M \leq C_{const,2}^M \cap C_{foc,2}^L \leq C_{const,2}^L$. If so, we can conclude that the true value of $f(\tilde{\alpha})$ is given by Equation (14), which implicitly determines $f_2(\tilde{\alpha})$. If the condition does not hold though, the true value of $f(\tilde{\alpha})$ is given by any other functions $f(\tilde{\alpha})$, as defined by Equation (13) or the expressions in Equations (A.1.1)-(A.1.6) in Appendix A.1. We repeat the analysis for a value of $f(\tilde{\alpha})$ obtained with Equation (13), $f_1(\tilde{\alpha})$, and test if the condition $C_{foc,2}^H > C_{const,2}^H \cap C_{foc,2}^M > C_{const,2}^M \cap C_{foc,2}^L > C_{const,2}^L$ holds. If the previous condition is not satisfied, we repeat the experiment with Equation (A.1.1), taking into account the implicit condition that the corresponding consumption paths must hold, and so on, successively up to the last value function. Thus, we obtain different portfolios weights and determine the corresponding function $f_*(\tilde{\alpha})$, for each portfolio weight.[15] We find that for 99.90% of the portfolio weights, the functions to be employed are either $f_1(\tilde{\alpha})$, or $f_2(\tilde{\alpha})$. Therefore, in most of the cases, $C_{foc,2}^H > C_{const,2}^H \cap C_{foc,2}^M > C_{const,2}^M \cap C_{foc,2}^L > C_{const,2}^L$ or $C_{foc,2}^H \leq C_{const,2}^H \cap C_{foc,2}^M \leq C_{const,2}^M \cap C_{foc,2}^L \leq C_{const,2}^L$. The difference among the different scenarios reflects the stock returns $(H, M, L)$; in a state of nature, when consumption (as obtained under the first-order condition or constraint) is higher than another form of consumption, this inequality persists for the remaining states of nature.

The next step is to find the portfolio weights among the set of portfolios defined by the grid of $\tilde{\alpha}s$ that maximizes $f_*(\tilde{\alpha})$. The optimal portfolio weight vector is denoted $\tilde{\alpha}^*$, and it gives the value function of the modified problem. After obtaining $\tilde{\alpha}^*$ and $f_*(\tilde{\alpha})$, we can easily derive the optimal consumption path in the different liquidity regimes and scenarios. Thus, the main point of this optimization problem is to maximize the function $f(\tilde{\alpha})$, which has been implicitly determined. The advantage of this methodology is the use of numerical methods to compute the corresponding $f(\tilde{\alpha})$ for each value of $\tilde{\alpha}$ and thus obtain the optimal portfolio $\tilde{\alpha}^*$. This methodology implies a substantial simplification in comparison with other numerical methods that rely on iterative procedures to approximate the value function.

---

[15] A detailed review of the different possibilities appears in Appendix A.1.

## 2.5. Calibration

In this section, we provide adequate ground for the values of the benchmark parameters we use in the portfolio problem optimization. The parameter $\beta$ is set equal to 0.95, in agreement with the values employed in macroeconomic literature, including Jofre-Bonet and Pesendorfer (2000) and Asiedu and Villamil (2000).

Previous literature has employed a wide variety of values for the relative risk aversion coefficient. According to Gollier (2001), it seems reasonable to assume a relative risk aversion coefficient ranging between 1 and 4. Flavin and Yamashita (2002), for instance, assume a coefficient of relative risk aversion equal to 2. We adopt a conservative approach and choose $\theta$ equal to 2 for all households in the sample to prevent the results being unduly influenced by high risk aversion.

The time interval is one month, which means that an order is executed one month after it is placed. Although this time interval may be too long for stocks, it is not particularly high for other assets, taking into account households' efforts to rebalance portfolios, and the time spent to do so. According to the model, the time interval is also the length of the liquidity shock.[16] A shorter interval may imply a very short length of time with regard to the illiquidity restrictions on selling the assets and a very short duration of the effective consumption period.

We assume that the liquidity shocks occur once every four years on average and so, in annualized terms $\mu = 0.25$. We also assume that the size of the shock is equal to 1.18 ($\gamma = 1.18$), so if there is liquidity shock then consumption during this shock gives the individual 1.4 times more utility than consuming during normal liquidity regimes (i.e., $\gamma^\theta$). To confirm the suitability of these parameters, we employ implied moments for the growth of final consumption expenditures of Spanish households. The standard deviation of annual consumption growth equals 2.79%, in line with the data reported by Campbell (2003) for France (2.9%) and Germany (2.43%). However, because we assume that portfolio rebalancing takes place monthly and our benchmark is the standard deviation of quarterly consumption growth equal to 3.2%, we calibrate $\mu$ and $\gamma$ to ensure that the standard deviation of the liquidity shock is compatible with our benchmark and the level of extra utility accords with common sense. Thus, a liquidity shock increases the utility of consumption, but an increment 100 or 1000 times higher than in periods without shock would conflict with common sense; the value proposed by Koren and Szeidl (2002), where $\gamma^\theta = 11^2 = 121$, thus seems too high). Setting $\mu = 0.25$ and $\gamma = 1.18$, the standard deviation of the liquidity shock is 3.25%, which is very similar to our benchmark value. We use quarterly consumption to calibrate the parameter $\gamma$, because Spain's monthly consumption data is not available. Moreover, the similar standard deviation observed in the annual and quarterly consumption growths indicates that the results will not be unduly sensitive to the use of annual or quarterly consumption. So, we can capture patterns observed in quarterly consumption that likely would be cancelled out in annual data. The annualized benchmark parameters employed in the analysis are as follows: $\beta = 0.95, \theta = 2, \Delta t = 1/12, \gamma = 1.18,$ and $\mu = 0.25$.

---

[16] Koren and Szeidl (2002) recognize that this point seems to create a potential weakness, but if a liquidity shock has a longer length than the waiting time before the trade takes place, the household can optimally trade to counteract continuing liquidity shock after the trade is executed. Thus, a longer liquidity shock should not change the optimal portfolio substantially.

We employ annual historical real returns from 1991 (one year prior to the first year for which we have information on homeowners in the EFF) to the date of the survey, or the year we set as the initial date in the optimization problem (2002 or 2005). Because we consider three different scenarios, we should set different returns for the financial assets, depending on the realized scenario. However, the complexity of the model makes it infeasible to associate three different scenarios with each of the considered assets. The riskiest asset is the stock index, so we include stock index returns depending on the state of nature. We further assume that bank time deposits, mortgage, and housing returns are defined by the average of the historical returns with no volatility, so the stocks are the only assets that present risk to the households' portfolios.[17] Panel A of Table 1 summarizes the descriptive statistics of the historical real returns. The standard deviation of the stocks is ten times larger than the standard deviation of the other assets, supporting our assertion that the main source of risk in the households' portfolio is stock returns' volatility. Moreover, with the exception of housing, these asset returns show a low correlation with stock returns.[18,19]

<Insert Table 1 here>

## 3. Data and Sample

The data employed in this study is based on the Spanish Survey of Household Finance (EFF), managed by the Bank of Spain. We focus on the information about the demographic characteristics and the financial and economic situation of Spanish households in the years 2002 and 2005. Additionally, the EFF offers certain retrospective information about housing (year, price of acquisition, and amount and conditions of the mortgage at the origination date).

---

[17] Appendix A.2 includes details about the estimation of the housing returns

[18] Housing and stocks engage in a negative relationship, such that the bursting of the dot-com bubble in the early 2000s coincided with the starting point of the process of rapid revalorization in housing prices, which persisted up to 2008. Nevertheless, we do not take into account the correlation between housing and stock returns for several reasons. First, the investment share in housing is not used in the optimization problem, so using different states of nature for the housing returns would not have any effect on the optimal investment in housing. Second, the volatility of housing returns is almost seven times lower than the volatility of stock returns, so the housing returns in the different scenarios should not change materially from the average return value.

[19] Bank time deposits and mortgages are highly correlated, though that correlation does not generate any problem with regard to portfolio weights because for these assets we assume a constant return through the states of nature.

Appendix A.2 includes a definition of the variables we constructed, such as housing returns, mortgage payments, the value of the desired housing, household financial constraints (wealth and income), the degree of financial sophistication, and consumption.

A key variable for defining borrowing constraints and the portion of wealth invested in housing is the housing value at the survey moment, as reported by the household. Thus, we eliminate data from households that bought their housing before 1992 because, the further from the survey date, the more subjective is the value that the household reveal. This step also guarantees that any retrospective imputation used to calculate, for example, the mortgage payment will offer at least minimum reliability. To reduce the heterogeneity that could influence portfolio choices, we discard data from households in which the head is older than 60 years of age. The households in our sample dwell in their desired housing, as we noted previously, because we consider housing a long-run asset whose purchase determines others investments. If we include households in a house whose value is conspicuously below that of their desired housing, their investments plans could be influenced by the desire to change houses, which would shift the main asset in Spanish households' portfolios. We also exclude data from households that have undertaken repairs to their housing that were valued at more than 50,000 Euros, to avoid deviations between the real housing market value and the subjectively considered value claimed by the households after the renovations. Moreover, if such a significant renovation took place, then it would be difficult to assume that the house was the one desired at the moment of the purchase.

The households in the EFF have different probabilities of entering the sample, so we associate each household with its corresponding selection weight, to ensure we obtain representative statistics of the population. We use the weights to calculate the descriptive statistics as well, such that the mean and median are averages across the corresponding values of the statistics in the five imputations, and the variances are obtained according to the user's guide indications.[20]

The final sample consists of 427 households for the first imputation of the 2002 survey. The EFF reports the representativeness (weights) of the households in the Spanish population, so we can translate this number of households into a value that indicates their representativeness in the total population. These weights also enable us to construct different estimates and their variances. The 427 households are representative of 1,249,355 Spanish households. We provide the main descriptive statistics in Panel B of Table 1. Housing accounts for the main portion of households' wealth, and it is worth noting the remarkable difference between the housing's purchase value and its market value in 2002, which suggests a growing asset appreciation and emphasizes the importance of including it as a determinant restriction when analyzing Spanish households' portfolios. Regarding the contributions to the total net wealth of different assets, the data shows the importance of housing; on average, it is 1.282 times net wealth. Households usually resort to external funding to buy housing, so the rate of the mortgage over the wealth is 31.9%. The proportion of net wealth invested in deposits is 3.1%, whereas stocks represent 0.6% of total net wealth.

---

[20] To make inferences we use five multiple imputed data sets. To obtain a point estimate such as a mean or median we use the average of the five estimates obtained in each of the five imputations that form the survey. The total variance for the estimate is the sum of the within-imputation sampling variance and 6/5 times the between-imputations variance. See Bover (2004) and Barceló (2006) for a detailed description of imputation in the EFF.

We also employ the EFF 2005 survey to investigate portfolio rebalancing for a total of 130 households (representative of 390,300 households) for which we have information from both the 2002 and the 2005 waves. Moreover, we employ 230 households (representative of 714,902 households) from the 2002 survey who bought housing between 1997 and 2001, as well as 323 households (representative of 1,169,080 households) from the 2005 survey who bought housing between 2000 and 2004.[21]

As for other data used in the subsequent estimations, the housing price index came from the Secretary of State for Housing. The variables related to the interest rates of bank time deposits and EURIBOR were obtained from the Statistical Bulletin of the Bank of Spain. To generalize as much as possible and to avoid using maturities or the particularities of the different deposits, the bank time deposit interest rate we use corresponds to the Spanish Federation of Savings Banks (CECA) passive reference rate. This conservative methodology implies a widely accepted interest rate. The property tax values were obtained from different city council web-pages.

## 4. Empirical Results

### 4.1. Optimal Portfolio Composition

In Table 2, we report the optimal portfolio composition in our baseline analysis $(\beta, \theta, \Delta t, \gamma, \mu) = (0.95, 2, 1/12, 1.18, 0.25)$. We estimate the optimal portfolio composition for every household and imputation and then, aggregate the individual portfolios for nine categories into which a given household may be classified as available in the EFF: age, education, degree of financial sophistication, economic sector, labor situation, sex, household income, net wealth, and type of financial restrictions. The baseline model assigns optimal weights of 1.6%, 4.9%, and -34.7% to stocks, bank time deposits, and mortgage, respectively; the actual weights are 0.6%, 3.1%, and -31.9%. We observe that the optimal and actual weights are close in absolute terms but not in relative terms. In relative terms, the optimal investments in stocks, deposits, and mortgage are 166%, 58%, and 9% higher than the actual investments. Statistical tests indicate that optimal investments in stocks and deposits are significantly higher than actual ones. Therefore, Spanish households tend to invest less in stocks and deposits than theory indicates they should.

---

[21] We only present the descriptive statistics for the EFF 2002 survey because these data constitute the core of this investigation; the EFF 2005 data merely extend our main results of the paper. However, the descriptive statistics for households in the EFF 2005 survey are available on request.

Next, we analyze the effect of several households' characteristics on their allocation decision to understand the sources of mismatch in the allocation choices. To conduct this analysis we stratify the results obtained for individual households according to different households' characteristics. Note that such characteristics are neither used in the portfolio choice decision nor in the calibration of the model.

### 4.1.1. Equity

The optimal investment in stocks varies considerably among different subgroups: from 0.2% (households constrained in wealth or in both wealth and income) to 7.4% (financially sophisticated households). The model posits a positive and linear relation between the optimal investment in stocks and particular categories, such as the educational level, the household net wealth, and the economic sector. This linear relationship also emerges in the actual portfolios. Lower education levels imply lower investments in stocks (0.7%); households headed by a person with a university education should invest the highest proportion (2.3%). This pattern is consistent with Haliassos and Bertaut's (1995) finding that education is important for overcoming barriers to stockholding erected by ignorance. The increasing effect of net wealth is in line with Campbell (2006), Carroll (2002), and King and Leape (1998) findings: wealthy households are willing to take greater risk in their portfolios. In our case, the optimal investment for the households with the lowest net wealth is 0.5% while for the households with the highest net wealth this share is equal to 4.2%. Regarding the economic sector, the optimal stock proportion for households headed by a person who works in the tertiary sector (1.7%) is higher than the shares for the cases in which the person of reference works in the secondary (1.6%) or primary (0.7%) sectors. This pattern is consistent with the one observed in the actual weights.

For groups defined by age, degree of financial sophistication, or household income, the model instead predict a non-linear relationship with stock investment. In the case of age, the optimal maximum investment in stocks should be by households headed by a person between the ages of 45 and 55 years (3.2%), and decreasing in the following age group. This result is not consistent with the actual data, which shows a linear relationship between stock ownership and age. For the degree of financial sophistication, the model posits maximum investments by highly sophisticated households (7.4%) and much lower shares for the less and mildly sophisticated households (2.2% and 1.4%, respectively) while the actual data show a linear positive relationship. With regards to household income, the model posits a non-linear relationship being the shares for the three groups of households with earnings lower than €60,000 much lower than the optimal share for households that earn more than 60,000€ (3.4%). In the actual data though, there is a linear and positive relationship between stock ownership and household income.

The gender of the head of the household has no significant effect in either the model or the actual data. However, the theoretical investment weights (1.5 - 1.6%) again are significantly higher than the observed (0.5 - 0.6%). For the labor situation, the theoretical pattern and the actual one are broadly consistent. The highest investment comes from the retired group, and the lowest is associated with the self-employed group. Bertaut and Starr-McCluer (2001) obtain similar results for U.S. data. This result likely reflects the idea that investing in one's own business makes a household reluctant to increase their risk exposure any further.

In many cases the optimal investment is significantly higher than the actual one (under-investment). By categories, we find that older/retired, highly sophisticated, unconstrained or constrained in wealth, and rich households, either in terms of income or net wealth; do not significantly deviate from their optimal investment path. On the contrary, employed, low income, young, and less sophisticated households tend to invest below the optimal level. The under-investment by less financially sophisticated households may imply that these households fail to invest in stocks or invest cautiously because they are aware that they lack the skills to invest efficiently (Calvet, Campbell, and Sodini, 2006). Finally, we observe that independently of the education level and sex, households tend to under-invest in stocks.

### 4.1.2. Bank Time Deposits

The investment in bank time deposits does not exhibit a clear pattern across different categories and also reveals much lower variation than stocks. The proportion invested in bank time deposits varies from 2.4% (head of household is retired) to 5.7% (head of household employed in the primary sector). In some cases, there is a negative relation between the optimal investment in bank time deposits and some demographic categories, such as age and economic sector. The actual proportions are almost constant with age (with a decrease in the 35 - 55 years age group) and non-linear with the sector. Household income and net wealth indicate a non-linear z-shape in their optimal weights, in contrast with the v-shaped pattern observed in actual data. According to the labor situation, households headed by a retired person should invest the lowest proportion in bank time deposits (2.4%), and the highest proportion of investment should correspond to self-employed households (5.5%). This behavior is mirrored in the actual weights: 1.9% and 4.6%, respectively. The optimal an actual investment shares are both relatively similar between men and women. Households unconstrained by either income or wealth should invest less than households with any type of constraint. On the contrary, the actual proportions show that unconstrained households invest the highest proportion (4.4%) in this category.

In almost all cases the optimal investment in deposits is significantly higher than the actual one but there are some exceptions. Thus, the richer and unconstrained households tend to invest optimally in deposits. The age seems to be another factor that leads to optimal investments in deposits given that the investments of the older and retired households are not significantly different from the optimal investment. Not surprisingly, investments by highly financially sophisticated households do not differ significantly from the optimum. The less financially sophisticated households, the ones with secondary education and the self-employed households also invest efficiently in deposits.

### 4.1.3. Mortgage and Housing

Mortgage and housing are closely interconnected such that the lower the proportion of housing, the lower the proportion of mortgage. Recall that we consider the investment in housing as an already committed investment; we optimize the weight of the mortgage in the household portfolio, conditional on the investment in housing. The proportion (optimal and actual) of housing and mortgage decreases with the age and education. We also observe a v-shaped pattern in both optimal and actual weights relative to the degree of sophistication, economic sector, and household income categories. The optimal mortgage proportions decrease with net wealth, whereas in the actual proportions, we find an initial decrease, followed by an increase among the wealthiest group, which is somewhat surprising.

Regarding employment, the model predicts that retired households should have the lowest levels of mortgage (-7.3%) and housing (100.8%), because older households should have paid for most of their housing, as it is consistent with the actual weights. Men should have higher mortgage levels than women (-35.5% and -32.7%, respectively), partially because housing represents a higher proportion of their net wealth (129% and 126.5%, respectively). The constraint in wealth is the dominant constraint in this setting. Thus, we find similar proportions in housing and mortgage for households constrained in their wealth (-74.6%) and also for those constrained in both their wealth and income (-75.3%).

The remarkable non-linear patterns observed for different groups and assets mainly reflect the non-linear pattern of housing shares across these groups (e.g., degree of financial sophistication category). Moreover, for the cases that reveal a linear pattern in housing shares and a non-linear pattern for a given asset for a given group or category, the differences within each group and asset are low, considering the levels of the investment shares in absolute terms. In all cases, the optimal and actual proportions invested in mortgages are statistically indistinguishable.

Considering the three types of financial assets at once, we conclude that the households headed by highly financially sophisticated, older, retired, richer, and unconstrained persons are the ones investing more efficiently. Given the investment in housing and the financial markets' performance, the remaining categories of households should have a slightly higher share of mortgage to be invested in equity and bank-time deposits to increase the households' liquidity and capacity to face liquidity shocks. These results could contribute to a better design and implementation of financial education programs that may help households with certain characteristics to achieve effectively more optimal investments.

<Insert Table 2 here>

## 4.2. Optimal Portfolio Rebalancing and Investment Decision of Recent Homeowners

From the optimal and actual portfolios in 2002, we extend the analysis using the EFF 2005 survey. We study the portfolio rebalancing, from 2002 to 2005, for a total of 130 households for which we have information from both surveys. Panel A of Table 3 reports the difference between the optimal portfolio for the baseline optimization problem and the actual portfolio, according to the EFF information for the 2002 and 2005 surveys.[22]

Because we still assume that housing value is given by households' decision, there is no difference between the optimal and the observed housing value relative to wealth. On average, households underinvest in stocks and bank time deposits. The under-investment is similar in both surveys (1.3%) for stocks, but decreases for deposits from 1.7% to 1.2%. Finally, there is scarce difference between observed and optimal investment in mortgages.[23]

Then, we employ 230 households (from the 2002 EFF wave) that bought the housing between 1997 and 2001 and 323 households (from the 2005 EFF wave) that bought the housing between 2000 and 2004 to investigate the differences in the portfolio choices derived from supposedly less restrictive banking practices around 2005. Results are reported in Panel B of Table 3. The average optimal investment in mortgages is slightly higher in 2005 than in 2002 (-54.1% and -49.3%, respectively), perhaps due to the increase in housing prices and somewhat less restrictive banking practices. The optimal investment in bank deposits is similar in 2002 (5%) and 2005 (5.2%), but the optimal investment in stocks in 2005 (1.6%) is more than 70% higher than in 2002 (0.9%).

<Insert Table 3 here>

## 5. Comparative Statistics Analysis

This section presents a comparative statics analysis based on alternative values for the calibrated parameters $(\beta, \gamma, \mu, \Delta t)$, for our assumptions about banking practices in terms of the amount of wealth and income required for households to receive a mortgage, and for the stock portfolio's return. We report the results in Table 4. Regarding the parameter $\beta$, we explained in Section 2 the calibration procedure and why 0.95 provides a benchmark value; here we employ two alternative discount factors: 0.975 and 0.99. By increasing the discount factor, we decrease the penalty for delaying consumption. The proportion invested in stocks slightly decreases with the discount factor, and with a higher discount factor, the proportion invested in deposits declines. Because larger $\beta s$ imply more concern about the future, households should be willing to decrease their investments in stocks and deposits, as well as their leverage, when $\beta$ varies from the baseline.

---

[22] Note that we are considering 130 households instead of the 427 households that composed the 2002 wave so results' for the year 2002 are not strictly comparable with the ones in Table 2.
[23] Because the mortgage shares are negative, the signs should be interpreted in the opposite way: when the difference between the optimal and the observed portfolio is negative, it implies an underinvestment.

In Table 4, we also report the comparative statics related to the size of the liquidity shock ($\gamma$). The baseline is 1.18, and the alternatives for the parameter $\gamma$ are 1.1 and 1.5. The higher the liquidity shock, the higher is the proportion of investments in deposits, likely because a stronger liquidity shock increases the deposit's utility and so, households prefer to maintain larger proportions of ready available "risk-free" assets to avoid the possible losses to stock investments if the scenario turns unfavorable. The investment in stocks does not change across the different sizes of the taste/liquidity shock. As the size of the shock increases, the mortgage share also increases. The mortgage could even be used as an additional source of liquidity to obtain cash that could be invested in deposits.[24]

In the baseline scenario, we calibrate the parameter $\mu$ such that it is equal to 0.25, which means that there is a liquidity shock every four years. We set other values for this parameter (0.15, 0.35) and find that, similarly to the outcomes for the parameter $\gamma$, the higher the probability of a liquidity shock, the higher is the proportion invested in deposits. The intuition also remains the same: both parameters are related to the liquidity shock, so the more probable the liquidity shock, the larger the proportions of "risk-free" assets households maintain to avoid possible losses due to an unfavorable stock scenario. The optimal share to be invested in stocks thus slightly decreases as the parameter $\mu$ increases. As occurs with the optimal investment in the mortgage when we analyze variations in parameter $\gamma$, the higher the probability of a shock, the higher is the optimal investment. This result stresses the role of the mortgage as a source of liquidity that gets invested in deposits whenever the loan-to-value ratio is below 0.8.

The length of the liquidity shock ($\Delta t$) is one month in the baseline case. For longer periods, the proportion invested in stocks increases and the proportion invested in deposits decreases. As the shock duration increases, households prefer to invest in assets with more risk but also higher expected returns. The mortgage share also decreases as the shock duration increases.

Finally, we evaluate the effect of bank practices (restrictions *ix* and *x* in Equation (3)) on the portfolio composition and mortgage demand in Spain. Mayordomo (2008) analyzes changes in banking sector practices related to the mortgage-granting process and finds that if the mortgage increased to constitute more than the 80% of housing value, housing demand would increase considerably. However, other changes in these practices, such as income requirements or variations in the mortgage interest rate, have a lower impact. Therefore, we estimate how households' decision would change if the financial institutions change their practices. Note that if we restrict the maximum loan-to-value ratio to 70%, the optimization is not possible for some households, namely those whose mortgage loan-to-value exceeds that figure. If the loan-to-value ratio decreases to 70%, some households could not have become homeowners. If we discard these households from the optimization problem, the share of housing decreases. In contrast, when the maximum loan-to-value ratio increases to 90%, some households for which the optimization problem could not be solved because their mortgage loan-to-value exceeded 80% enter the analysis, which increases the overall housing share. Moreover, this less restrictive banking policy increases the weight of the mortgage in the households' portfolios. However, the optimal share of stocks and deposits does not change materially for the different loan-to-value ratios.

---

[24] According to the low variation in the standard deviation of the annual and quarterly consumption growth, we use slightly higher and lower values than 3.25%, to acknowledge that the standard deviation of the monthly consumption may be slightly higher or lower than the previous value. For a standard deviation in the monthly consumption growth equal to 3%, we find that the parameter γ should be equal to 1.16; for a standard deviation equal to 3.5%, the same parameter should be equal to 1.20. We thus confirm the low variation in the parameter γ for such levels of volatility. We find very similar results using any of the three values for parameter γ.

Change in the income requirements for mortgage payments does not cause material variations in the optimal portfolio composition. This result in line with Mayordomo's (2008) finding that income requirements have a negligible impact on a household's housing purchase decision. We extend this result to other financial assets in the household portfolio.

Finally, instead of using the previous percentiles of the stock returns' distribution, which define the states of nature, we now consider less extreme scenarios. Concretely, we define the favorable state when the annual return of the stock index exceeds the 66th percentile of its distribution in 2002 (30.02%), the intermediate scenario when returns fall between the 33rd and 66th percentiles (16.86%), and the unfavorable scenario when the returns are below the 33rd percentile (-11.23%). Thus, the return under a favorable, intermediate, and unfavorable scenario is 30.02, 16.86, and -11.23%, respectively. Under these scenarios, we find a very weak shift from stocks to deposits in the optimal portfolio being the new optimal shares in deposits and stocks 1.4% and 5.1%, respectively. Additionally, we consider an international instead of a domestic stock index (EuroStoxx50) and define the stock index returns as in the baseline specification on the basis of the 25th percentile (-8.34%), the average return in the range 25th-75th percentiles (8.96%), and the 75th percentile (30.23%) percentiles. The returns defining these percentiles are close to the ones employed for the 33rd and 66th percentiles of the IBEX35 returns and so, the results are identical.

<Insert Table 4 here>

## 6. Conclusions

Two approaches can be followed to model optimal portfolio choices. On the one hand, one may include housing as an additional asset in the standard portfolio choice problem. On the other, one may consider housing as the primary asset that determines the composition of the rest of the portfolio. In the latter case, the housing investment represents a decision already made by households, so estimates of the optimal portfolio must be conditional on the housing value. We adopt the latter approach in this paper and restrict the optimization problem to the households living in their desired housing to reduce the risk that they would move or rent the house in the short-run.

In the theoretical part of the paper, we present an optimization problem with four different financial assets: stocks, deposits (our proxy for the risk-free asset), mortgage, and housing. We apply this model in the empirical section to a unique set of micro data from the Bank of Spain's EFF. We present estimates of optimal portfolios for different individual households and for different groups of households, defined according to their demographic characteristics. Finally, after having estimated the optimal portfolio, we compare it with households' actual portfolio choices and study the factors affecting the deviations between both portfolios.

Our baseline results show that, given the actual proportion invested on average by Spanish households in housing, they invest significantly less in stocks and deposits than theory indicate they should. The optimal investment in mortgage is slightly higher than the observed one but the difference between them is not statistically significant. We also find a positive relationship between the optimal proportion invested in stocks and the households' education level and net wealth. The proportion of housing and the optimal share of mortgage decreases with age, education level, and household net wealth. Finally, the optimal amount to be invested in bank time deposits is around 4-5% and nearly constant across demographics groups. Considering the three types of financial assets at once, we conclude that the households headed by highly financially sophisticated, older, retired, richer, and unconstrained persons are the ones investing more efficiently. The remaining categories of households could have a slightly higher share of mortgage to be invested in equity and/or bank-time deposits to increase the households' liquidity and capacity to face liquidity shocks.

## Appendix A.1 Consumption Schedules and Implicit Expression of Value Functions

In this appendix, we define the different consumption schedules and the corresponding implicit expression of the value functions, $f_i(\tilde{\alpha})$ for $i = 3,\ldots,8$, depending on the liquidity shock realization and the state of nature:[25]

- If $\left(C_{foc,2}^H > C_{constr,2}^H\right) \cap \left(C_{foc,2}^M > C_{constr,2}^M\right) \cap \left(C_{foc,2}^L > C_{constr,2}^L\right)$, then $f(\tilde{\alpha}) = f_1(\tilde{\alpha})$.

$$C_{constr,2}^i = \left(\alpha_1 + \alpha_2(1-fee) + r_p^i\right)W + L \quad for\ i = H, M, L \qquad (\text{A.1.a})$$

- If $\left(C_{foc,2}^H \leq C_{constr,2}^H\right) \cap \left(C_{foc,2}^M \leq C_{constr,2}^M\right) \cap \left(C_{foc,2}^L \leq C_{constr,2}^L\right)$, then $f(\tilde{\alpha}) = f_2(\tilde{\alpha})$.

$$C_{foc,2}^i = \frac{f_2(\tilde{\alpha})\beta^{-1/\theta}(R_p^i W + L)}{1 + \beta^{-1/\theta} f_2(\tilde{\alpha})} \quad for\ i = H, M, L \qquad (\text{A.1.b})$$

---

[25] The value function $f(\tilde{\alpha})$ should be obtained as the root of the corresponding equations.

- If $\left(C^H_{foc,2} > C^H_{constr,2}\right) \cap \left(C^M_{foc,2} > C^M_{constr,2}\right) \cap \left(C^{L,3}_{foc,2} \leq C^L_{constr,2}\right)$, then $f(\tilde{\alpha}) = f_3(\tilde{\alpha})$.

$$1 = [a(\mu + (1-\mu)(\lambda_1 + \lambda_2))] + f_3(\tilde{\alpha})^\theta \gamma^\theta \mu[\lambda_1 b + \lambda_2 c + \lambda_3 d] + f_3(\tilde{\alpha})^\theta (1- \mu)[\lambda_1 b + \lambda_2 c] + (1-\mu)\beta\lambda_3 R_p^{L\,1-\theta}\left(1 + \beta^{\frac{-1}{\theta}} f_3(\tilde{\alpha})\right)^\theta \quad (A.1.1)$$

$$C^{L,3}_{foc,2} = \left[f_3(\tilde{\alpha})\beta^{\frac{-1}{\theta}}(R_p^L W + L)\right] / \left(1 + \beta^{\frac{-1}{\theta}} f_3(\tilde{\alpha})\right) \quad (A.1.c)$$

$$C^i_{constr,2} = (\alpha_1 + \alpha_2(1-fee) + r_p^i)W + L \quad for\ i = H, M$$

where
$$a = (1 - \tilde{\alpha}_1 - \tilde{\alpha}_2(1-fee))^{1-\theta}\beta$$

$$b = \left(\tilde{\alpha}_1 + \tilde{\alpha}_2(1-fee) + r_p^H - \tilde{\alpha}_2 r_2 + \frac{L}{W}\right)^{1-\theta}$$

$$c = \left(\tilde{\alpha}_1 + \tilde{\alpha}_2(1-fee) + r_p^M - \tilde{\alpha}_2 r_2 + \frac{L}{W}\right)^{1-\theta}$$

$$d = \left(\tilde{\alpha}_1 + \tilde{\alpha}_2(1-fee) + r_p^L - \tilde{\alpha}_2 r_2 + \frac{L}{W}\right)^{1-\theta}$$

$$\lambda_3 = 1 - \lambda_1 - \lambda_2$$

- If $\left(C^H_{foc,2} > C^H_{constr,2}\right) \cap \left(C^{M,4}_{foc,2} \leq C^M_{constr,2}\right) \cap \left(C^L_{foc,2} > C^L_{constr,2}\right)$, then $f(\tilde{\alpha}) = f_4(\tilde{\alpha})$.

$$1 = [a(\mu + (1-\mu)(\lambda_1 + \lambda_3))] + f_4(\tilde{\alpha})^\theta \gamma^\theta \mu[\lambda_1 b + \lambda_2 c + \lambda_3 d] + f_4(\tilde{\alpha})^\theta (1- \mu)[\lambda_1 b + \lambda_3 d] + (1-\mu)\beta\lambda_2 R_p^{M\,1-\theta}\left(1 + \beta^{\frac{-1}{\theta}} f_4(\tilde{\alpha})\right)^\theta \quad (A.1.2)$$

$$C^{M,4}_{foc,2} = \left[f_4(\tilde{\alpha})\beta^{\frac{-1}{\theta}}(R_p^M W + L)\right] / \left(1 + \beta^{\frac{-1}{\theta}} f_4(\tilde{\alpha})\right) \quad (A.1.d)$$

$$C^i_{constr,2} = (\alpha_1 + \alpha_2(1-fee) + r_p^i)W + L \quad for\ i = H, L$$

- If $\left(C^{H,5}_{foc,2} \leq C^H_{constr,2}\right) \cap \left(C^M_{foc,2} > C^M_{constr,2}\right) \cap \left(C^L_{foc,2} > C^L_{constr,2}\right)$, then $f(\tilde{\alpha}) = f_5(\tilde{\alpha})$.

$$1 = [a(\mu + (1-\mu)(\lambda_2 + \lambda_3))] + f_5(\tilde{\alpha})^\theta \gamma^\theta \mu[\lambda_1 b + \lambda_2 c + \lambda_3 d] + f_5(\tilde{\alpha})^\theta (1- \mu)[\lambda_2 c + \lambda_3 d] + (1-\mu)\beta\lambda_1 R_p^{H\,1-\theta}\left(1 + \beta^{\frac{-1}{\theta}} f_5(\tilde{\alpha})\right)^\theta \quad (A.1.3)$$

$$C^{H,5}_{foc,2} = \left[f_5(\tilde{\alpha})\beta^{\frac{-1}{\theta}}(R_p^H W + L)\right] / \left(1 + \beta^{\frac{-1}{\theta}} f_5(\tilde{\alpha})\right) \quad (A.1.e)$$

$$C^i_{constr,2} = (\alpha_1 + \alpha_2(1-fee) + r_p^i)W + L \quad for\ i = M, L$$

- If $\left(C^H_{foc,2} > C^H_{constr,2}\right) \cap \left(C^{M,6}_{foc,2} \leq C^M_{constr,2}\right) \cap \left(C^{L,6}_{foc,2} \leq C^L_{constr,2}\right)$, then $f(\tilde{\alpha}) = f_6(\tilde{\alpha})$.

$$1 = [a(\mu + (1-\mu)\lambda_1)] + f_6(\tilde{\alpha})^\theta \gamma^\theta \mu[\lambda_1 b + \lambda_2 c + \lambda_3 d] + f_6(\tilde{\alpha})^\theta (1-\mu)\lambda_1 b +$$
$$(1-\mu)\beta[\lambda_2 R_p^{M\,1-\theta} + \lambda_3 R_p^{L\,1-\theta}]\left(1 + \beta^{\frac{-1}{\theta}} f_6(\tilde{\alpha})\right)^\theta \quad (A.1.4)$$

$$C_{foc,2}^{i,6} = \left[f_6(\tilde{\alpha})\beta^{\frac{-1}{\theta}}(R_p^i W + L)\right] \Big/ \left(1 + \beta^{\frac{-1}{\theta}} f_6(\tilde{\alpha})\right) \text{ for } i = M, L \quad (A.1.f)$$

$$C_{constr,2}^H = (\alpha_1 + \alpha_2(1-fee) + r_p^H)W + L$$

- If $(C_{foc,2}^{H,7} \le C_{constr,2}^H) \cap (C_{foc,2}^M > C_{constr,2}^M) \cap (C_{foc,2}^{L,7} \le C_{constr,2}^L)$, then $f(\tilde{\alpha}) = f_7(\tilde{\alpha})$.

$$1 = [a(\mu + (1-\mu)\lambda_2)] + f_7(\tilde{\alpha})^\theta \gamma^\theta \mu[\lambda_1 b + \lambda_2 c + \lambda_3 d] + f_7(\tilde{\alpha})^\theta (1-\mu)\lambda_2 c +$$
$$(1-\mu)\beta[\lambda_1 R_p^{H\,1-\theta} + \lambda_3 R_p^{L\,1-\theta}]\left(1 + \beta^{\frac{-1}{\theta}} f_7(\tilde{\alpha})\right)^\theta \quad (A.1.5)$$

$$C_{foc,2}^{i,7} = \left[f_7(\tilde{\alpha})\beta^{\frac{-1}{\theta}}(R_p^i W + L)\right] \Big/ \left(1 + \beta^{\frac{-1}{\theta}} f_7(\tilde{\alpha})\right) \text{ for } i = H, L \quad (A.1.g)$$

$$C_{constr,2}^M = (\alpha_1 + \alpha_2(1-fee) + r_p^M)W + L$$

- If $(C_{foc,2}^{H,8} \le C_{constr,2}^H) \cap (C_{foc,2}^{M,8} \le C_{constr,2}^M) \cap (C_{foc,2}^L > C_{constr,2}^L)$, then $f(\tilde{\alpha}) = f_8(\tilde{\alpha})$.

$$1 = [a(\mu + (1-\mu)\lambda_3)] + f_8(\tilde{\alpha})^\theta \gamma^\theta \mu[\lambda_1 b + \lambda_2 c + \lambda_3 d] + f_8(\tilde{\alpha})^\theta (1-\mu)\lambda_3 d +$$
$$(1-\mu)\beta[\lambda_1 R_p^{H\,1-\theta} + \lambda_2 R_p^{M\,1-\theta}]\left(1 + \beta^{\frac{-1}{\theta}} f_8(\tilde{\alpha})\right)^\theta \quad (A.1.6)$$

$$C_{foc,2}^{i,8} = \left[f_8(\tilde{\alpha})\beta^{\frac{-1}{\theta}}(R_p^i W + L)\right] \Big/ \left(1 + \beta^{\frac{-1}{\theta}} f_8(\tilde{\alpha})\right) \text{ for } i = H, M \quad (A.1.h)$$

$$C_{constr,2}^L = (\alpha_1 + \alpha_2(1-fee) + r_p^L)W + L$$

Ultimate consumption is determined by the grid of $\alpha$s that maximizes $f_*(\tilde{\alpha})$. When we have maximized $f(\tilde{\alpha})$, we can easily derive the optimal consumption path in the different liquidity regimes and scenarios.

## Appendix A.2 Definition of Variables

### A.2.1. Housing Returns

The estimation of the housing returns is based on the estimation of the user cost of homeowners provided by Diaz and Luengo-Prado (2008). The housing returns can be defined as:

$$hr_{it} = \frac{q_{it}(1 - \delta + \tau) - q_{it-1} + rent_{it}}{q_{it-1}} \quad (A.2.a)$$

where $hr_{i,t}$ represents the housing returns per square meter for a given household i at period $t$, and $q_{i,t}$ and $q_{i,t-1}$ are the housing prices per square meter at periods *t* and *t-1*, respectively. The parameter $\delta$ represents housing's rate of depreciation, which is set to 0.043 as in Díaz and Luengo-Prado (2008). The parameters $\tau$ is the property tax in Spain, equal to 0.0068. This tax comes into effect at a local jurisdiction level and we employ an average value that reflects the tax for Madrid and Barcelona. Finally, $rent_{it}$ reflects the income per square meter that the household would have received if it had rented the housing.

### A.2.2. Mortgage Payments

The annual payments were obtained from the remaining amount of the loan principal at the survey year (2002), according to a French amortization system and using either fixed or variable interest rates, depending on the loan's characteristics.

### A.2.3. Value of the desired housing

To estimate the desired value, we employ a subsample formed by (i) financially unconstrained households, (ii) households in which the head of household is between 25 and 60 years old, and (iii) households that bought their housing within the previous 5 years. We employ a generalized Tobit model to estimate the desired value. The censure is set according to the expression $V_i \geq Z_i$ being $V_i$ the housing purchase price, and $Z_i \equiv \min\left(\frac{NW_i}{1-0.8}, \frac{0.33Y_i}{0.8r}\right)$ the minimum of the limit values set when a household is constrained in wealth or rent, equivalent to constraints (*ix*) and (*x*) in Equation (3). Under the Tobit specification, we regress the housing purchase price on a group of variables whose values correspond with the year the housing was bought: permanent income, net wealth, the user cost of the housing, age, and different demographic characteristics of the head of the household. The estimated coefficients enable us to predict the desired value by a wider sample of households. The equation estimated with the subsample is used to infer or predict the desired value for a wider sample of households that includes constrained members.

### A.2.4.1. Household wealth financial constraints

We define a limit value ($V_i^W$) that indicates if a given household *i* is constrained or unconstrained in wealth as:

$$V_i^W = \left(\frac{NW_i}{1 - 0.8}\right) \qquad (\text{A.2.b})$$

where $NW_i$ is household *i*'s net wealth the year the housing was bought. The limit value $V_i^W$ is related to the restriction (*ix*) in Section 2, which implies that the initial payment for the housing purchase must be at least 20% of the housing price. The value 0.8 is the maximum portion of the mortgage with respect to the housing purchase price that banks offer borrowers. We consider a given household *i* constrained in wealth whenever the desired housing value is higher than the value limit $V_i^W$.

*A.2.4.2. Household income financial constraints*

We define a limit value ($V_i^Y$) that indicates if a given household $i$ is constrained or unconstrained in income as:

$$V_i^Y = \frac{0.33 Y_i}{0.8 r} \qquad (A.2.c)$$

where $Y_i$ is household $i$'s total annual income the year it bought housing, and $r$ is the mortgage rate. The limit value $V_i^Y$ relates to restriction (*x*) in Section 2, which implies that mortgage payments must be lower than 33% of household income. The value 0.33 reflects the maximum portion of the mortgage payment with respect to household income. We consider a given household $i$ constrained in income whenever the desired housing value is higher than the value limit $V_i^Y$.

*A.2.5. Grade of financial sophistication*

To construct a proxy of the financial knowledge of a given household, we consider seven different groups of financial instruments or actions that may indicate strong financial knowledge: realization of electronic payments; investment in options, futures, swaps or other derivatives; use of credit cards; use of checks; use of direct billing or direct deposit; use of telephone banking; and use of Internet banking. For a given household, each action equals 1 if the household uses it and 0 otherwise. We consider a household *less sophisticated* when the sum of these values is less than 3. A household is a *midly sophisticated* investor if the sum of these values is between 4 and 5, inclusive. Finally, a given household is *highly sophisticated* if the sum is 6 or 7.

*A.2.6. Consumption*

The consumption variable represents households' expenditures, including food, but excluding durable goods, housing rentals or other property costs, mortgage payments, insurance, housing alterations, and housing maintenance costs.

Table 1: Descriptive Statistics

This table reports the descriptive statistics of the main databases in the sample. Panel A contains information about the historical real asset returns calculated using annual information from 1991 to 2002. It reports the mean ($E[R]$) and the standard deviation ($SD[R]$) of the annual returns which are deflated by the appropriate price index. For the stocks, we present the mean returns under a favorable ($R_H$); an intermediate ($R_M$), and an unfavorable scenario ($R_L$). It also contains the correlation matrix for the different pairs of returns. Panel B reports the mean, median and standard deviation of the households' demographic characteristics, financial assets and housing. The statistics refer to the EFF 2002 survey wave.

Panel A

|  | Stocks | | | Bank Time Deposits | Housing | Mortgage |
|---|---|---|---|---|---|---|
|  | L | M | H |  |  |  |
| E[R] | -13.87 | 16.86 | 37.77 | 1.15 | 3.66 | 4.53 |
| SD[R] |  | 28.96 |  | 1.11 | 4.32 | 2.18 |

Correlation Matrix

|  | Stocks | Bank Time Deposits | Housing | Mortgage |
|---|---|---|---|---|
| Stocks | 1 |  |  |  |
| Bank Time Deposits | 0.217 | 1 |  |  |
| Housing | -0.715 | -0.631 | 1 |  |
| Mortgage | 0.204 | 0.971 | -0.673 | 1 |

Panel B

|  | Mean | Median | Std. Dev. |
|---|---|---|---|
| Age of the head of household | 39.979 | 38.000 | 8.897 |
| Head of household has primary education (1 if yes, 0 otherwise) | 0.374 | 0.000 | 0.484 |
| Head of household has secondary education (1 if yes, 0 otherwise) | 0.378 | 0.000 | 0.485 |
| Head of household has university education (1 if yes, 0 otherwise) | 0.248 | 0.000 | 0.432 |
| Less sophisticated investor (1 if yes, 0 otherwise) | 0.166 | 0.000 | 0.372 |
| Midly sophisticated investor (1 if yes, 0 otherwise) | 0.832 | 1.000 | 0.374 |
| Highly sophisticated investors (1 if yes, 0 otherwise) | 0.002 | 0.000 | 0.039 |
| Head of household works in the primary sector (1 if yes, 0 otherwise) | 0.058 | 0.000 | 0.234 |
| Head of household works in the secondary sector (1 if yes, 0 otherwise) | 0.292 | 0.000 | 0.455 |
| Head of household works in the tertiary sector (1 if yes, 0 otherwise) | 0.606 | 1.000 | 0.489 |
| Head of household is employed (1 if yes, 0 otherwise) | 0.687 | 1.000 | 0.464 |
| Head of household is self-employed (1 if yes, 0 otherwise) | 0.124 | 0.000 | 0.329 |
| Head of household is retired (1 if yes, 0 otherwise) | 0.013 | 0.000 | 0.115 |
| Head of household is unemployed or non-active (1 if yes, 0 otherwise) | 0.186 | 0.000 | 0.389 |
| Sex of the head of household (1 if men, 0 women) | 0.702 | 1.000 | 0.457 |
| Annual Income (€) | 38,693 | 29,916 | 33,509 |
| Annual Labor Income (€) | 32,208 | 26,728 | 25,230 |
| Annual Consumption (€) | 12,302 | 10,800 | 7,813 |
| Net Wealth (€) | 159,542 | 131,460 | 108,845 |
| Stocks (€) | 902 | 0 | 16,073 |
| Deposits (€) | 5,035 | 2,151 | 11,254 |
| Mortgage (€) | -50,948 | -25,139 | 33,702 |
| Housing (€) | 204,552 | 153,193 | 100,586 |
| Portion of net wealth invested in stocks (%) | 0.006 | 0.000 | 0.026 |
| Portion of net wealth invested in deposits (%) | 0.031 | 0.016 | 0.048 |
| Portion of net wealth invested in mortgage (%) | -0.319 | -0.191 | 0.381 |
| Portion of net wealth invested in housing (%) | 1.282 | 1.165 | 0.386 |
| Housing purchase price (€) | 90,741 | 83,440 | 60,755 |
| Year of housing purchase | 1,997 | 1,997 | 2.497 |
| Household constrained in wealth (1 if yes, 0 otherwise) | 0.281 | 0.000 | 0.449 |
| Household constrained in income (1 if yes, 0 otherwise) | 0.901 | 1.000 | 0.297 |
| Household constrained in wealth and income (1 if yes, 0 otherwise) | 0.263 | 0.000 | 0.440 |
| Household unconstrained (1 if yes, 0 otherwise) | 0.080 | 0.000 | 0.271 |

Table 2: Optimal Portfolio Choice Baseline Results

This table reports the optimal portfolio (i.e., optimal proportion of stocks, deposits, mortgage and housing) of the baseline optimization problem ($\beta = 0.95, \theta = 2, \Delta t = 1/12, \gamma = 1.18,$ and $\mu = 0.25$), and the actual portfolio attending to the EFF information for seven subgroups referred to the head of household: age, education, degree of financial sophistication, economic sector of employment, labor situation, sex, income, net wealth, and type of financial constraints that household faces. The first group of columns reports the estimation of the optimal portfolio while the second group of columns reports the actual portfolio. The symbols ** and * indicate whether the optimal shares are significantly different from the observed shares at 1 and 5% significance levels, respectively.

| Subgroup | Optimal Portfolio | | | | Observed Portfolio | | | |
|---|---|---|---|---|---|---|---|---|
| | Stocks | Bank Time Deposits | Mortgage | Housing | Stocks | Bank Time Deposits | Mortgage | Housing |
| All Individuals | | | | | | | | |
| All Individuals | 0.016** | 0.049** | -0.347 | 1.282 | 0.006 | 0.031 | -0.319 | 1.282 |
| Age | | | | | | | | |
| Less than 35 | 0.006* | 0.052** | -0.452 | 1.394 | 0.002 | 0.035 | -0.430 | 1.394 |
| Between 35 and 45 | 0.012** | 0.048** | -0.384 | 1.323 | 0.004 | 0.025 | -0.353 | 1.323 |
| Between 45 and 55 | 0.032** | 0.046 | -0.181 | 1.103 | 0.010 | 0.035 | -0.147 | 1.103 |
| More than 55 | 0.029 | 0.045 | -0.125 | 1.051 | 0.019 | 0.035 | -0.104 | 1.051 |
| Education | | | | | | | | |
| Primary Education | 0.007** | 0.051** | -0.387 | 1.329 | 0.001 | 0.025 | -0.356 | 1.329 |
| Secondary Education | 0.019* | 0.045 | -0.329 | 1.265 | 0.006 | 0.035 | -0.305 | 1.265 |
| University Education | 0.023* | 0.051** | -0.312 | 1.238 | 0.012 | 0.035 | -0.285 | 1.238 |
| Degree of Financial Sophistication | | | | | | | | |
| Less Sophisticated | 0.022* | 0.044 | -0.248 | 1.182 | 0.003 | 0.030 | -0.213 | 1.182 |
| Midly Sophisticated | 0.014** | 0.050** | -0.367 | 1.303 | 0.006 | 0.031 | -0.341 | 1.303 |
| Highly Sophisticated | 0.074 | 0.047 | -0.052 | 0.931 | 0.090 | 0.063 | -0.086 | 0.931 |
| Economic Sector | | | | | | | | |
| Primary Sector | 0.007 | 0.057** | -0.390 | 1.327 | 0.001 | 0.027 | -0.353 | 1.327 |
| Secondary Sector | 0.016 | 0.050* | -0.408 | 1.342 | 0.006 | 0.033 | -0.380 | 1.342 |
| Tertiary Sector | 0.017** | 0.047** | -0.323 | 1.259 | 0.006 | 0.032 | -0.297 | 1.259 |
| Labor Situation | | | | | | | | |
| Employed | 0.016** | 0.047** | -0.355 | 1.291 | 0.006 | 0.029 | -0.327 | 1.291 |
| Self-Employed | 0.012* | 0.055 | -0.320 | 1.252 | 0.003 | 0.046 | -0.301 | 1.252 |
| Retired | 0.041 | 0.024 | -0.073 | 1.008 | 0.014 | 0.019 | -0.041 | 1.008 |
| Unemployed/Non-Active | 0.015 | 0.052** | -0.344 | 1.277 | 0.005 | 0.031 | -0.313 | 1.277 |
| Sex | | | | | | | | |
| Men | 0.015** | 0.050** | -0.355 | 1.290 | 0.005 | 0.032 | -0.326 | 1.290 |
| Women | 0.016** | 0.046** | -0.327 | 1.265 | 0.006 | 0.030 | -0.302 | 1.265 |
| Income | | | | | | | | |
| Less than 20,000€ | 0.015* | 0.049** | -0.328 | 1.264 | 0.002 | 0.031 | -0.297 | 1.264 |
| Between 20,000 and 40,000€ | 0.012** | 0.048** | -0.387 | 1.327 | 0.002 | 0.027 | -0.355 | 1.327 |
| Between 40,000 and 60,000€ | 0.011 | 0.050** | -0.358 | 1.297 | 0.008 | 0.034 | -0.339 | 1.297 |
| More than 60,000€ | 0.034 | 0.047 | -0.261 | 1.180 | 0.020 | 0.040 | -0.241 | 1.180 |
| Net Wealth | | | | | | | | |
| Less than 100,000€ | 0.005 | 0.053** | -0.664 | 1.606 | 0.002 | 0.027 | -0.635 | 1.606 |
| Between 100,000 and 200,000€ | 0.015** | 0.046** | -0.249 | 1.187 | 0.003 | 0.031 | -0.222 | 1.187 |
| Between 200,000 and 300,000€ | 0.023* | 0.049** | -0.119 | 1.047 | 0.010 | 0.030 | -0.087 | 1.047 |
| More than 300,000€ | 0.042 | 0.048 | -0.114 | 1.024 | 0.023 | 0.047 | -0.093 | 1.024 |
| Type of Financial Restrictions | | | | | | | | |
| Constrained in Wealth | 0.002 | 0.054* | -0.746 | 1.690 | 0.003 | 0.038 | -0.731 | 1.690 |
| Constrained in Income | 0.014** | 0.049** | -0.352 | 1.289 | 0.005 | 0.030 | -0.324 | 1.289 |
| Constrained in Wealth and Income | 0.002 | 0.056* | -0.753 | 1.697 | 0.003 | 0.038 | -0.737 | 1.697 |
| Unconstrained | 0.036 | 0.048 | -0.215 | 1.131 | 0.020 | 0.044 | -0.194 | 1.131 |

Table 3: Portfolio Rebalancing and Investment by Recent Homeowner

This table reports information about the optimal portfolio rebalancing and the investment decision by recent homeowners. Panel A contains the deviation between the optimal portfolio (i.e., optimal proportion of stocks, deposits, mortgage and housing) for the baseline optimization problem ($\beta = 0.95, \theta = 2, \Delta t = 1/12, \gamma = 1.18, \text{and } \mu = 0.25$) and the actual portfolio attending to the EFF information for the 2002 and 2005 surveys. The households employed in this analysis are the ones for which we have information from both surveys. The first group of columns reports the deviation between the optimal and the actual portfolios in 2002 while the second group of columns reports the equivalent result in 2005. The symbols ** and * indicate whether the deviation between the optimal and observed shares is significantly significant at 1 and 5% significance levels, respectively. Panel B reports the optimal portfolio for the recent homeowners. The first group of columns reports the optimal portfolio using the information of the housing buyers between 1997 and 2001 in the 2002 survey. The second group of columns reports the optimal portfolio using the information of the housing buyers between 2000 and 2004 in the 2005 survey.

| | Panel A | | | | | | | |
|---|---|---|---|---|---|---|---|---|
| | Deviation from the optimal in 2002 | | | | Deviation from the optimal in 2005 | | | |
| Subgroup | Stocks | Bank Time Deposits | Mortgage | Housing | Stocks | Bank Time Deposits | Mortgage | Housing |
| All Individuals | 0.013** | 0.017** | -0.030 | 0.000 | 0.013** | 0.012** | -0.025 | 0.000 |
| | Panel B | | | | | | | |
| | Optimal Portfolio for the Recent Homeowners in 2002 | | | | Optimal Portfolio for the Recent Homeowners in 2005 | | | |
| Subgroup | Stocks | Bank Time Deposits | Mortgage | Housing | Stocks | Bank Time Deposits | Mortgage | Housing |
| All Individuals | 0.009 | 0.050 | -0.493 | 1.434 | 0.016 | 0.052 | -0.541 | 1.473 |

Table 4: Optimal Portfolio Choice under other Parameter's Values

This table reports the optimal portfolio choice under different parameter's values. The results presented in this table correspond to the average of all the individuals' portfolio choices using the subsample of 427 households from the EFF 2002 data. In the first column we show the different values employed for each parameter (beta, gamma, mu, time, theta, loan to value, income over mortgage payment and state of nature) while in the second column we show the optimal investment in the corresponding financial asset (stocks, deposits, mortgage, and housing). In the case of the state of nature we vary the probability of the state of nature using the IBEX35 and EuroStoxx50 as benchmark.

| Parameters Values | Stocks | Bank Time Deposits | Mortgage | Housing |
|---|---|---|---|---|
| Beta | | | | |
| Baseline ($\beta$=0.95) | 0.0156 | 0.0487 | -0.3467 | 1.2823 |
| $\beta$=0.975 | 0.0147 | 0.0381 | -0.3352 | 1.2824 |
| $\beta$=0.99 | 0.0152 | 0.0300 | -0.3270 | 1.2820 |
| Gamma | | | | |
| $\gamma = 1.1$ | 0.0150 | 0.0481 | -0.3450 | 1.2820 |
| Baseline ($\gamma = 1.22$) | 0.0150 | 0.0500 | -0.3467 | 1.2823 |
| $\gamma = 1.5$ | 0.0153 | 0.0617 | -0.3590 | 1.2823 |
| Mu | | | | |
| $\mu = 0.15$ | 0.0164 | 0.0396 | -0.3383 | 1.2823 |
| Baseline ($\mu = 0.25$) | 0.0156 | 0.0487 | -0.3467 | 1.2823 |
| $\mu = 0.35$ | 0.0142 | 0.0574 | -0.3539 | 1.2823 |
| Time (Shock Duration in annual terms) | | | | |
| $\Delta t = 1/24$ | 0.0140 | 0.0520 | -0.3480 | 1.2820 |
| Baseline ($\Delta t = 1/12$) | 0.0156 | 0.0487 | -0.3467 | 1.2823 |
| $\Delta t = 1/4$ | 0.0245 | 0.0192 | -0.3257 | 1.2820 |
| Loan-to-Value Ratio | | | | |
| LTV = 70% | 0.0158 | 0.0480 | -0.3334 | 1.2696 |
| Baseline (LTV = 80%) | 0.0156 | 0.0487 | -0.3467 | 1.2823 |
| LTV = 90% | 0.0155 | 0.0493 | -0.3523 | 1.2875 |
| Income over Mortgage Payment | | | | |
| IMP = 25% | 0.0156 | 0.0487 | -0.3467 | 1.2823 |
| Baseline (IMP = 33%) | 0.0156 | 0.0487 | -0.3467 | 1.2823 |
| IMP = 40% | 0.0165 | 0.0490 | -0.3470 | 1.2820 |
| Stock Index Returns (States of nature) | | | | |
| Ibex35 (25th - 75th percentiles) | 0.0156 | 0.0487 | -0.3467 | 1.2823 |
| Ibex35 (33rd - 66th percentiles) | 0.0129 | 0.0514 | -0.3466 | 1.2823 |
| EuroStoxx50 (25th - 75th percentiles) | 0.0142 | 0.0507 | -0.3472 | 1.2823 |

Figure 1: Model Timing

This figure illustrates the timing for asset trade, consumption, and interest earned.

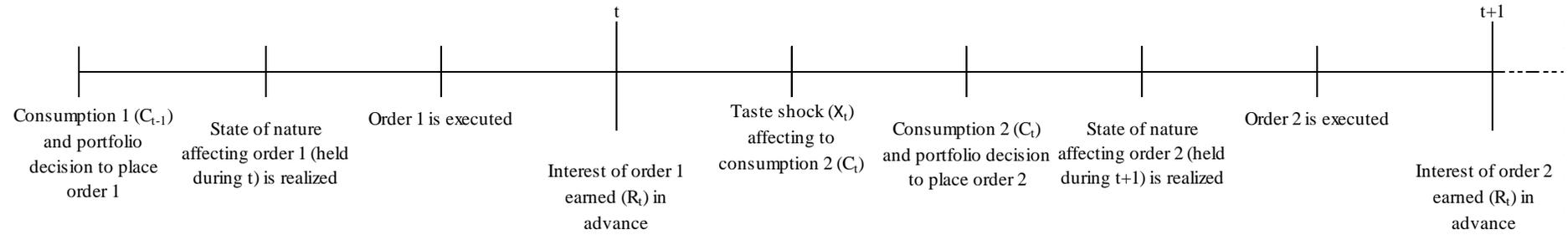